\documentclass[prl,letterpaper,twocolumn]{revtex4}
\usepackage{amsmath,amssymb}
\usepackage{graphicx,color}
\newcommand{\eref}[1]{~(\ref{#1})} % *** FIX all eqn refs to use this style.
\newcommand{\pws}{\ensuremath{P_\text{ws}}}
\newcommand{\dws}{\ensuremath{\Delta_\text{ws}}}

\newcommand \be  {\begin{equation}}
\newcommand \beq {\begin{equation}}
\newcommand \bea {\begin{eqnarray} \nonumber }
\newcommand \ee  {\end{equation}}
\newcommand \eeq {\end{equation}}
\newcommand \eea {\end{eqnarray}}
\newcommand{\beqa}{\begin{eqnarray}}
\newcommand{\eeqa}{\end{eqnarray}}

\newcommand{\nn}{\nonumber\\}

\newcommand{\calv}{\ensuremath{\mathcal{V}}}
\newcommand{\calvc}{\ensuremath{{\mathcal{V}}(c)}}

\newcommand{\bol}[1]{{\boldsymbol{#1}}}
\newcommand{\give}[1]{\overset{#1}{\longrightarrow}}

\newcommand{\rme}{\text{e}}

\newcommand{\xp}{\ensuremath{\bol{x}_p}}
\newcommand{\pp}{\ensuremath{\bol{p}_p}}

\begin{document}

\title{Exchange of stability as a function of system size in a nonequilibrium system}
\author{Sorin T\u{a}nase-Nicola and David K. Lubensky}

\affiliation{Department of Physics, University of Michigan, Ann Arbor MI  48109-1040, USA}

\begin{abstract}
In equilibrium systems with short-ranged interactions, the relative stability of different thermodynamic states generally does not depend on system size (as long as this size is larger than the interaction range).  Here, we use a large deviations approach to show that, in contrast, different states \textit{can} exchange stability as system size is varied in a driven, bistable reaction-diffusion system.  This striking effect is related to a shift from a spatially uniform to a nonuniform transition state and should generically be possible in a wide range of nonequilibrium physical and biological systems.

\end{abstract}

\maketitle

The statistical physics of nonequilibrium systems has proven to be an enduring source of unexpected and intriguing phenomena.  Historically, studies of driven systems have tended to focus on the thermodynamic limit of infinite system size.  In recent years, however, experimental advances have made it possible to study everything from micromagnets and nanotubes to living cells at the mesoscopic scale, and this in turn has led to a growing theoretical interest in finite-sized stochastic systems.  A major problem is to understand the relative stability of, and transition rates among, different (meta)stable states; this issue has been studied extensively for some equilibrium, detailed-balance-obeying models~\cite{Faris82,Maier01,Burki08}.  Here, we consider the same question in a simple model of a far-from-equilibrium, driven chemical system and show that far richer behavior is possible when detailed balance is violated.  In particular, we find that---unlike in an equilibrium system---the system's two stable states can exchange stability as the system size increases.  This reversal is related to competition between a homogeneous and a spatially-varying transition state~\cite{Maier01,Meerson11}, and we expect it generically to be possible whenever there is no underlying Boltzmann distribution to ensure that the relative stability is independent of the transition path.  Similar results may thus apply to a wide range of systems of interest in condensed matter, chemical, and biological physics.

In what follows, we start by introducing the Schl\"{o}gl model of a bistable chemical system and formulating its mean-field rate equations in terms of a \textit{deterministic potential} \calvc.  The minima of this potential correspond to two locally stable states, and one might naively imagine that the state with the lower \calv\ is the more stable of the two.  Indeed, in a spatially extended system with diffusive transport, the mean-field equations have front solutions in which the state with lower \calv\ always invades that with higher \calv.  In a well-stirred system, in contrast, only noise-induced transitions between states are possible, and the mean-field description provides no information about relative stability.  Surprisingly, in this case, an analysis that accounts for the fundamentally stochastic nature of chemical reactions shows that, in certain parameter ranges, the state with \textit{higher} \calv\ is found with greater probability~\cite{Hanggi84,Doering07}.  To reconcile these two seemingly contradictory predictions, we consider a model including both diffusion and fluctuations.  We first show numerically that the two states exchange stability as the system's spatial size grows.  We then use a semiclassical approach to explain this dependence.  By placing bounds on the action that determines the transition rate between states, we are able to show that the relative stability must follow the deterministic potential for large enough systems, where the transition occurs through nucleation followed by deterministic front motion, while the well-stirred result applies for small enough systems.
 
{\bf The bistable Schl\"{o}gl model}~\cite{Schlogl72} consists of the chemical reactions
\beq
A \give{k_0} X, \, X \give{k_1} A, \,
2 X   +B \give{k_2} 3 X, \, 3 X \give{k_3} 2 X + B \;,
\label{eq:sch}
\eeq
where the concentrations of $A$ and $B$ are held constant. The corresponding 
mean-field rate equation for the concentration $c$ of $X$ in a well-stirred system is
\beq
\dot c =\left(k_0+ k_2  c^2\right) -\left(k_3 c^3+ k_1 c\right) =-{\cal V}'(c) \; ,
%\dot c =\left(k_0+ k_2  c^2\right) -\left(k_3 c^3+ k_1 c\right) =-\frac{d{\cal V}(c)}{dc} \; ;
\label{eq:chemdyn}
\eeq
where we have absorbed the concentrations of $A$ and $B$ into the rates $k_0$ and 
$k_2$, respectively, and the prime denotes a derivative.  This equation defines the deterministic potential $\calv(c)$.  If we move away from the well-stirred limit and let $c$ depend on a spatial coordinate $z$, a diffusion term must be added, and Eq.\eref{eq:chemdyn}\ generalizes to
\beq
\partial_t c =-\calv'(c) +{\cal{D}}\partial^2_z c \; .
\label{eq:rd}
\eeq

For appropriate choices of the $k_i$, ${\cal V}(c)$ has one local maximum $c_s$ and two local 
minima $c_1$ and $c_2$, with $c_1 < c_2$,  corresponding  to two (meta)stable  states.  Eq.\eref{eq:chemdyn}\ is invariant with respect to a simultaneous rescaling of concentrations and of rate constants, a fact we can exhibit explicitly by introducing a typical concentration scale $c_0$ of the same order as $c_1$ and $c_2$ and writing the $k_i$ as $k_i = \lambda_i (c_0)^{1-i}$, where the $\lambda_i$ have dimensions of inverse time.  For fixed $\lambda_i$, the dynamics of $x = c/c_0$ is independent of $c_0$.  

The existence of two (meta)stable states invites the question of their relative stability.  One expects that, when noise is properly taken into account, $c$ will be found with high probability near $c_1$ or $c_2$, and one might guess that $\calv(c)$ determines which of the two is more probable (i.e. stable) and which less probable.  Indeed, in the limit of infinite system size, Eq.\eref{eq:rd} admits traveling front solutions of the form $c(z,t) = f(z - v t)$ in which the state with lower \calv\ expands into the one with higher \calv~\cite{Faris82,Meerson11,Khain11}.

Eq.\eref{eq:rd}, however, is only a mean-field approximation to a more realistic model that accounts for the random nature of the molecular collisions that lead to chemical reactions.  To incorporate both these intrinsic stochastic effects and diffusion, we use a mesoscopic compartment model.  In general, a $d$-dimensional reaction 
vessel is partitioned into $M$ elementary compartments of linear size $h$ and volume $V = h^d$, and each molecule
 can jump between neighboring compartments but can react only with other molecules in 
the same compartment.  The stationary probability $P(n^1,n^2,\ldots,n^i,\ldots)$ to find $n^i$ molecules in 
compartment $i$ then satisfies
\bea
&&\sum_i\left[W_+(n^i-1)P(.,n^i-1,.)-W_+(n^i)P(.,n^i,.)\right]+\nonumber\\
&&\sum_i\left[W_-(n^i+1)P(.,n^i+1,.)-W_-(n^i)P(.,n^i,.)\right]+\nonumber\\
&&D\sum_{(i,j)}\left[ (n^i+1) P(.,n^i+1, n^j-1,.) - n^i P(.,n^i, n^j,.)\right]\nonumber\\
&&=0 \; ,
\label{eq:readiff}
\eea
where $D$ is the jump rate between compartments and the last sum is taken over neighboring compartments $(i,j)$. 
The reaction rates are 
\bea
W_+(n)& = &k_0 V +(k_2/V) n (n-1) \;, \nn 
W_-(n)& = & k_1 n + (k_3/ V^2) n (n-1)(n-2) \; .
\label{eq:rates}
\eea
In general, this model allows for barriers that slow diffusion between compartments, but we are primarily interested in using it as an approximate description of a spatially continuous system.  One can show that the average behavior of the stochastic compartment model\eref{eq:readiff} approaches that of the continuum reaction-diffusion equation\eref{eq:rd}, with $\mathcal{D} = h^2 D$, when
$D t_r = \mathcal{D} t_r/h^2 \gg 1$ and $c_0 V \gg 1$ \cite{Elf04, Isaacson06}.
  Here $t_{r} \sim 1/\lambda_i$ is a typical time between reactions of an individual molecule. 
The first 
inequality ensures that individual compartments are well mixed; while this could be accomplished by decreasing $h$ at fixed $\mathcal{D}$, the second inequality demands that this not be done at the expense of having very few molecules in each compartment.

\begin{figure}
\includegraphics[width=\columnwidth]{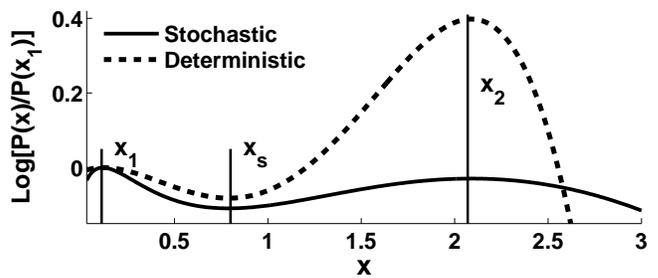}
\caption{Logarithms of the probability distribution $\pws(x)$ in the well stirred model and of the deterministic potential $\calv(c_0 x)$~\protect\cite{fig-paras}. Both quantities are scaled by their values at $x_1$.
 \label{fig:stochdim}}
\end{figure}

Although there is no general analytical expression for the stationary distribution satisfying Eq.\eref{eq:readiff}, one can be found when $M = 1$ and the entire reaction volume is well stirred.  In this case the total number of molecules $n$ is distributed according to
\beq
\pws(n)=K \prod_{j=1}^{n-1}\frac{W_+(j-1)}{W_-(j)} \; ,
%\pws(n)=K \prod_{j=1}^{n-1} W_+(j-1)/W_-(j) \; ,
\eeq
where $K$ is a normalization constant.  Below, we will be particularly interested in asymptotic results in the limit that the typical number of particles per compartment $\Omega \equiv c_0 V$ becomes large, and it is thus useful to rewrite \pws\ in terms of $x = c/c_0 = n/\Omega$ as~\cite{Dykman94} 
\beq
%\pws(x \Omega) = K(x;\Omega) \rme^{-\Omega S(x)} \; ,
\pws(x) = K(x;\Omega) \rme^{-\Omega S_\text{ws}(x)} \; ,
\label{eq:pws}
\eeq
where $K(x_1;\Omega)/K(x_2;\Omega)$ is bounded as $\Omega \rightarrow \infty$ for every $x_1,\, x_2>0$, so that $\exp[-\Omega S_\text{ws}(x)]$ determines the dominant large $\Omega$ contribution to any probability ratio.

The action $S_\text{ws}(x)$ is defined by 
\beq
dS_\text{ws}(x)/dx = \ln\left[w_-(x)/w_+(x)\right] \; ,
\label{eq:wss}
\eeq
where $w_+(x)=\lambda_0+\lambda_2 x^2$ and 
$w_-(x)=\lambda_1 x + \lambda_3 x^3$.  It is clear from this equation that the extrema of $S_\text{ws}(x)$ and of $\calv(c_0 x)$ occur at the same values of $x$.  The two functions, however, can otherwise be very different.  Indeed, for the parameters used in Fig.~\ref{fig:stochdim}~\cite{fig-paras}, ${\cal V}(c_2)-{\cal V}(c_1)$ and $\dws = S_\text{ws}(x_2) - S_\text{ws}(x_1)$ (where the extrema $x_i = c_i/c_0$) have opposite signs.  Since Eq.\eref{eq:pws}\ implies that it is the sign of \dws\ that determines which of the two states is more probable, the deterministic potential does not reliably predict relative stability in the well-stirred case.

\begin{figure}
\includegraphics[width=\columnwidth]{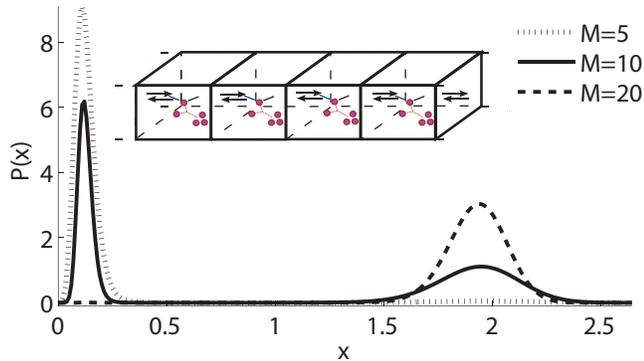}
\caption{The distribution of the average scaled concentration $x=\sum n_i/{(M \Omega)}$ for different $M$ and $V = 20$~\protect\cite{fig-paras}.  \textit{Inset}: schematic of the compartment model, with molecules allowed to hop between compartments and to react within each compartment.
 \label{fig:sim}}
\end{figure}

{\bf Semiclassical approach.}  In a well-stirred system, no transitions between states can occur without fluctuations.  In contrast, the kinetics\eref{eq:rd}\ with diffusion does allow for front-driven transitions between states even in the mean-field limit.  This suggests that, unlike in the well-stirred system, there may be cases in spatially extended systems where the deterministic potential \calvc\ {\em does} in fact determine relative stability.  Direct simulations of the fully stochastic compartment model (using a kinetic Monte Carlo algorithm with separate treatment of reaction and diffusion steps~\cite{Elf04}) demonstrate that this intuition is correct.  Indeed, Fig. \ref{fig:sim} shows that the two states can exchange stability as $M$ is increased; for small $M$, their relative stability is the same as in the well stirred limit, but for larger $M$, the state with lower \calv\ regains the upper hand.  As we now show, this stability inversion can be understood within a semiclassical approximation.

Just as in the well-stirred case, the stationary distribution $P$ satisfying Eq.\eref{eq:readiff} can be written for large $\Omega$ as
\beq
P({x^1,\dots, x^M})=K_M({x^1, \dots, x^M};\Omega)\rme^{-\Omega S({x^1, \dots, x^M})} \; ,
\label{eq:ldtm}
\eeq
where $x^i=\frac{n^i}{\Omega}$, and $\Omega = c_0 V$ remains the typical number of particles per compartment~\cite{Dykman94}.  For well-mixed systems, the eikonal approximation\eref{eq:ldtm}\ is often thought of as a large volume approximation, but here the requirement that the compartment model approximate a continuous reaction-diffusion system constrains the volume of each compartment, and it is more useful to think of making $\Omega$ large by letting $c_0 \rightarrow \infty$ with the $\lambda_i$, $D$, and $V$ fixed.

It is known \cite{Heymann08, Dykman94}  that the function $S(\bol{x})$ can be expressed as the minimal action
\bea
S(\bol{x}) - S(\bol{x}_i) = \min_{\bol{x}(t)}\max_{\bol{p}(t)} \left\{ \int_{\bol{x}_i}^{\bol{x}} \!\!dt\, \left[\bol{p} \dot{\bol{x}} - H(\bol{p},\bol{x})\right]\right\}\;, \label{eq:act} \\
H(\bol{p},\bol{x}) = \nonumber \\%\hspace*{6cm}  \\
\sum_i(\rme^{p^i}-1)\left[w_+(x^i)-w_-(x^i)\rme^{-p^i}+ 
D\Delta_i (\bol{x} \rme^{-\bol{p}})\right] , %\nonumber
\label{eq:defs}
\eea
where $\bol{x}_i$ is a stable stationary point of the deterministic kinetics, $\bol{x}$ a point in its basin of attraction, and $\Delta_i$ the appropriate discrete Laplacian centered on compartment $i$. The extrema of $S(\bol{x})$ correspond to the fixed 
points of the dynamical system 
\beq
\dot x^i=w_+(x^i)-w_-(x^i)+ D\Delta_i (\bol{x})
\label{eq:dyn}
\eeq
and have the same stability~\cite{Dykman94}. 

In the continuum limit $h \rightarrow 0$, Eq.\eref{eq:dyn}\ becomes a rescaled version of Eq.\eref{eq:rd}, whose stationary states are well-characterized~\cite{Maier01,Buttiker81,Scholl86,Hui00}.  It is reasonable to expect (see Supplemental Material [SM]) that the stationary states of the discretized version have similar properties for small enough $h$.  Then, with reflecting or periodic boundary conditions, only the uniform states $\bol{x_1}=(x_1,\dots,x_1)$ and $\bol{x_2}=(x_2,\dots,x_2)$ are stable, and there is a unique (up to symmetries) fixed point $\bol{x}_s$ of\eref{eq:dyn}\ with only one unstable direction.  This saddle is used to define the function $S(\bol{x})$  everywhere, i.e. to fix the value of $S(\bol{x}_i)$ and thus of the stability index
$\Delta$ through a matching procedure  \cite{Heymann08, Dykman94} :
\bea
\Delta=S(\bol{x}_2)-S(\bol{x}_1)=\Delta S_{1,s}-\Delta S_{2,s},\nn
\Delta S_{1,s}=
\min_{\bol{x}(t)}\max_{\bol{p}(t)}  \int_{\bol{x}_1}^{\bol{x}_s} \!\!dt\,\left[\bol{p} \dot{\bol{x}} - H(\bol{p},\bol{x})\right],\nn
\Delta S_{2,s}=
\min_{ \bol{x}(t)}\max_{\bol{p}(t)} \int_{\bol{x}_2}^{\bol{x}_s} \!\!dt\,\left[\bol{p} \dot{\bol{x}} - H(\bol{p},\bol{x})\right] \; .
\label{eq:defd}
\eea
Both $\Delta S_{i,s}$ are non-negative~\cite{Dykman94}.  

We now focus on the regime where (as in Fig.~\ref{fig:stochdim}) $\calv(c_2)-\calv(c_1) < 0$, but $\dws > 0$, and study the sign of $\Delta$, and through it the relative stability of the two uniform states.  This sign strongly depends on the saddle $\bol{x}_s$, which, in the continuum limit $h \rightarrow 0$, is spatially uniform for a small enough system but becomes nonuniform at a critical linear system size~\cite{Faris82,Maier01,Meerson11}.  In the limit of large system size, the saddle profile is close to $x_1$ 
everywhere except in a localized region whose size remains constant as the system size grows.  This form reflects that fact that the deterministic traveling waves favor the state $x_2$, so that only a small nucleus is required to initiate a transition from uniform $x_1$ to uniform $x_2$.  Similarly, for our compartment model, the stationary states and optimal trajectories in Eq.\eref{eq:defd} are uniform for $M$ less than some $M_c$, and one can show that $\Delta=M \dws$.  The stability is then that of a well
stirred system, independent of $D$. For $M>M_c$, solving the double optimization problem\eref{eq:defd}\
is extremely difficult.  One can, however, verify numerically (SM, Fig.~\ref{fig:Saddle}) that $\bol{x}_s$ has the same shape as in the continuum limit, and thus deviates appreciably from $x_1$ only in a small region.  Using this fact, one can derive bounds on $\Delta$.

In particular, it is not hard to see that $\Delta S_{1,s}$, which describes the difficulty of reaching $\bol{x}_s$ from $\bol{x}_1$, remains smaller than a fixed constant, independent of $M$. This bound is simply given by a particular trajectory solving the first maximization problem in
Eqs.\eref{eq:defd}. As $\bol{x}_s$, and thus the trajectory, differ from $\bol{x}_1$ only in a small region, the total action is finite, even for infinite $M$ (see SM).

\begin{figure}
\includegraphics[width=\columnwidth]{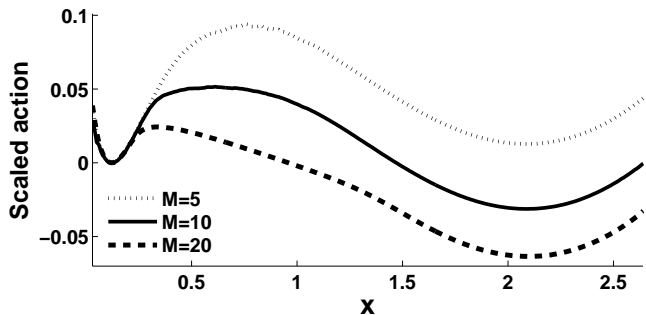}
\caption{The scaled action $-S/M$ for different numbers of compartments.  The plotted values represent the logarithm of the probability
of an average concentration $x = \sum_i x^i/M$ in Monte Carlo simulations~\protect\cite{Elf04}, extrapolated to $\Omega=\infty$.  Curves are shifted such that 
$S(x_1)=0$ in order to clearly indicate the stability exchange.  Note that the peaks in the curves move towards $x_1$ as $M$ grows, reflecting the fact that most of the saddle profile is near $x_1$ for large $M$.}
\label{fig:Action}
\end{figure}

Similar reasoning indicates that $\Delta S_{2,s}$ should grow linearly with $M$.  Indeed, in the limit $M \rightarrow \infty$, the region where $\bol{x}_s$ differs from $\bol{x}_1$ becomes negligible, and one can focus on an optimal trajectory that must take the system from $x_2$ to $x_1$ on an essentially infinite domain.  Physically, one expects that such an optimal trajectory should correspond to a front solution of the variational equations traveling with constant speed.  
Each successive compartment then makes the same contribution to the action, and one has $\Delta S_{2,s}(M) \simeq M \Delta_0$ for some $\Delta_0$.  (See also SM.)  Moreover, it is clear that for any integers $M$ and $l$, $\Delta S_{2,s}(M) \geq \Delta S_{2,s}(M l)/l$, because an optimal trajectory for a system of size $M$ can be mirrored $l$ times to create a suboptimal trajectory for a system of size $M l$.  Thus, the asymptotic large $M$ behavior $\Delta S_{2,s}(M) \simeq M \Delta_0$ implies that $\Delta S_{2,s}(M) \geq M \Delta_0$ even for finite $M$.  Since $\Delta S_{1,s}$ remains finite as $M$ increases while $\Delta S_{2,s}$ grows without bound, $\Delta = \Delta S_{1,s} - \Delta S_{2,s}$ must change sign at some $M > M_c$, and the two states exchange stability as the system size increases (Fig.~\ref{fig:Action}).

{\bf Discussion.} The large deviations approach that we have used to show that an exchange of stability must occur is formally valid in the limit that $\Omega \rightarrow \infty$ at fixed compartment volume $V$ and number of compartments $M$.  This order of limits is important.  In particular, because the prefactor $K_M$ in Eq.\eref{eq:ldtm} depends on $M$, it can overwhelm the exponential factor if $M$ tends toward infinity at fixed $\Omega$; thus, our calculations describe transitions between (meta)stable states in a finite-sized, mesoscopic system but cannot be used to study the phase transition in the Schl\"{o}gl model defined in the limit of infinite system size~\cite{Grassberger82}.  Similarly, our results hold in the limit of a large number of reacting particles and thus describe a distinct phenomenon from the exchange in the most probable state observed in a well stirred system with a finite (small) particle number~\cite{Ebeling79} or the recently-described noise-induced reversal of front propagation direction~\cite{Khain11}.  Our calculations also go considerably beyond a previous study that examined transition rates very near a bifurcation in the Schl\"{o}gl model but did not consider relative stability~\cite{Meerson11}.  Finally, although the exchange of stability discussed here could not happen if the transition state did not become non-uniform at $M = M_c$ (leading to a non-analyticity in $\Delta$)~\cite{Maier01, Burki08}, the actual exchange occurs at some $M > M_c$ and need not coincide with any further bifurcations of the transition state.

Although we have presented our results for a specific model of a bistable chemical system, the system-size-dependent relative stability that we describe is far more general and should in principle be possible in models describing everything from pattern formation~\cite{Bisang98} to ecological population dynamics~\cite{Meerson11} or the complex reaction networks present in living cells~\cite{Munoz-Garcia10}.  All that is absolutely required is multistability and violation of detailed balance.  In some cases, more complex stability diagrams are likely to be possible.  For example, if we no longer insist that the compartment model studied here be a good approximation to a continuum system and thus allow ourselves to vary $D$ arbitrarily, we expect a reentrant exchange of stability:  As $D \rightarrow \infty$, the entire system should be well mixed, while each individual compartment behaves as a separate well mixed vessel as $D \rightarrow 0$~\cite{Berglund07a}.  In either limit, the relative probability of the two states should depend on \dws, while for intermediate values of $D$ the sign of $\Delta$ may differ from that of \dws.  These and similar effects should be accessible within the same formalism employed here.

\textbf{Acknowledgments.}  We are grateful to Charlie Doering, Baruch Meerson, and Len Sander for helpful conversations.  This work was funded in part by NSF grant DMR1056456.

%\bibliographystyle{apsrev}
%\bibliography{biblio}

\begin{thebibliography}{22}
\bibitem{fig-paras} For all figures, the parameters are $ k_0=.2,\, k_1=2,\, k_2=3,\, k_3=1$, and $D = 1$, in units in which $c_0$ and $\lambda_3$ have been set to unity.
\end{thebibliography}

%%%%%%%%%%%%%%%%%%% BEGIN BIBLIOGRAPHY %%%%%%%%%%%%%%%%%%%%%%%%

%%%%%%%%%%%%%%%%%%%%%%%%%%% END BIBLIOGRAPHY %%%%%%%%%%%%%%%%%%%%%%%%%%%%%%%%

\clearpage

\section{Supplemental Material}

As in the latter part of the main text, we assume here that the parameters are such that $\calv(c_2) < \calv(c_1)$ but $\dws > 0$, so that the concentration $c_2$ is favored in the mean-field limit while $c_1$ is more stable for a well-stirred system.

\subsection{The stationary states}

The stationary states of Eq. (\ref{eq:rd}) satisfy
\beq
\mathcal{D} \partial_z^2 c = \calv'(c) 
\label{eq:ham}
\eeq
and represent trajectories of a Hamiltonian system in the classical potential $-\calvc$, where the
spatial variable $z$ plays the role of time.  In addition to the constant trajectories $c(z)=c_1,c_s,c_2$, nonuniform trajectories are also possible for large enough system lengths.

These trajectories represent one or more (half)oscillations around the
local minimum $c_s$ of $-\cal{V}$.  Intuitively each swing of these trajectories (or every full oscillation for periodic boundary conditions) can be independently destabilized, making the number of swings equal to the number of unstable directions of the corresponding stationary state. Therefore, when non-constant stationary states are possible, the saddle with only one unstable direction corresponds to a trajectory with a half (full) oscillation for reflecting (periodic) boundary conditions~\cite{Maier01, Buttiker81, Scholl86, Hui00}.

\begin{figure}
\includegraphics[width=3.5in]{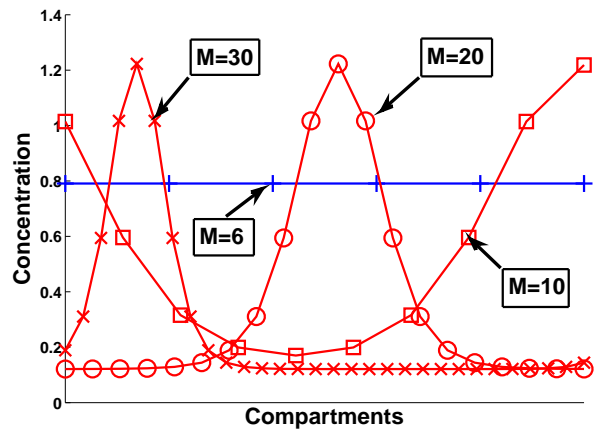}
\caption{
The saddle profile $\bol{x}_s$ representing a fixed point of Eq.~(\protect\ref{eq:dyn}) for periodic boundary conditions and different numbers $M$ of compartments. (The horizontal axis is rescaled so that profiles for different $M$ fit on the same graph.) For $M \leq M_c = 6$ the saddle profile is uniform.  As the number of compartments is increased for $M > M_c$, the profile converges to a form where most of the profile, except for a critical nucleus, is close to the metastable state $c_1$.}
\label{fig:Saddle}
\end{figure}

Fig.~\ref{fig:Saddle} shows numerically obtained saddle profiles for the discretized model of Eq.~(\ref{eq:dyn}).  A similar structure is obtained as in the continuum limit.  In particular, for large $M$ the region where the concentration deviates from $c_1$ remains the same size even as $M$ increases.  (Saddle profiles were obtained by solving the coupled algebraic equations defining the stationary state by standard methods, using the continuum solution as an initial guess, then verifying numerically that the linearized dynamics about the stationary profiles obtained in this manner had only a single positive eigenvalue.)

\subsection{Upper bound on $\Delta S_{1,s}$}
	
In order to construct an upper bound on the quantity 
\beq
\Delta S_{1,s}=\min_{\bol{x}(t)}\max_{\bol{p}(t)} \int_{\bol{x}_1}^{\bol{x}_s}\!dt\,\, \left[\bol{p} \dot{\bol{x}} - H(\bol{p},\bol{x})\right] \; ,
\eeq
it is enough to find a particular trajectory $\bol{x}_p(t)$ between $\bol{x}_1$ and $\bol{x}_s$ and then to evaluate
\beq
S[\bol{x}_p(t)]=\max_{\bol{p}(t)} \int_{\bol{x}_1}^{\bol{x}_s} \!dt\,\, \left[\bol{p} \dot{\bol{x}}_p - H(\bol{p},\bol{x}_p)\right]
\eeq
for that trajectory.  For a given choice of $\bol{x}_p$, the maximum in this equation is attained for momenta $\bol{p}_p$ satisfying
%\beq
%L=\bol{p} \dot{\bol{x}}_p - H(\bol{p},\bol{x}_p)
%\eeq
%satisfy
\beq
0=\dot x_p^i-\left.\frac{\partial H}{\partial p^i}\right|_{(\bol{x}_p,\bol{p}_p)}.
\label{eq:maxeq}
\eeq
Because the function
\bea
H(\bol{p},\bol{x})= \nonumber \\
\sum_i(\rme^{p^i}-1)\left[w_+(x^i)-w_-(x^i)\rme^{-p^i}+ 
D\Delta_i (\bol{x} \rme^{-\bol{p}})\right]
\eea
is convex, Eq.\eref{eq:maxeq} determines, for each $\bol{x}_p$, a unique $\bol{p}_p$ where a global maximum over all possible $\bol{p}(t)$ is attained.

We can take advantage of this uniqueness property to specify our particular trajectories \xp\ and \pp\ simultaneously.  For future convenience, we rewrite the momenta in terms of a (so far unknown) function $\bol{x}_*(t)$ as $p_p^i=\log{\left(\frac{x_p^i}{x^i_*}\right)}$.  Knowing that $H(\bol{p},\bol{x}) = 0$ along the true extremal trajectory~\cite{Dykman94}, we then define our particular trajectory by demanding that each term in the sum
 \bea
H(\bol{x}_p,\bol{p}_p)=\nonumber \\
\sum_i(\rme^{p_p^i}-1)\left[w_+(x_p^i)-w_-(x_p^i) x_*^i/x_p^i+ 
D\Delta_i (\bol{x}_*)\right]\;,
\eea
separately vanish:
\beq
0=w_+(x_p^i)-\frac{w_-(x_p^i)}{x_p^i} x_*^i+ D\Delta_i (\bol{x}_* )
\label{eq:spec}
\eeq
for each $i$.  These $M$ equations, together with the $M$ Hamilton's equations\eref{eq:maxeq}, determine \xp\ and \pp.  In particular, we can rewrite\eref{eq:maxeq} explicitly as a set of dynamical equations for the $x_p^i$:
\bea
\dot x_p^i=w_+(x_p^i) x_p^i/x_*^i - w_-(x_p^i) x_*^i/x_p^i + \nonumber \\
D (x_*^{i+1} x_p^i/x_*^i-x_p^{i+1} x_*^i/x_*^{i+1})+\nonumber \\
D (x_*^{i-1} x_p^i/x_*^i-x_p^{i-1} x_*^i/x_*^{i-1})\nonumber \; ,
\label{eq:xpdyn}
\eea
where Eq.\eref{eq:spec} implicitly determines the $x_*^i$ at each time.  It is not hard to verify that these dynamical equations for $\xp(t)$ share the same fixed points as the original mean-field kinetic equations for the concentrations $\bol{x}(t)$; indeed, for $M < M_c$, where the concentrations remain uniform along the optimal trajectory, $\xp(t)$ coincides with the true optimal trajectory.  In particular, one can easily check that the uniform state $x_p^i = x_*^i = x_1$ is a fixed point of Eq.\eref{eq:xpdyn} for any $M$.

We can now use the particular solution $(\pp,\xp)$ to show that $\Delta S_{1,s}$, remains finite even as $M \rightarrow \infty$.  First note that $H(\pp,\xp) = 0$ by construction.  Thus, our bound on $\Delta S_{1,s}$ reduces to
\beq
\Delta S_{1,s} \leq \int_{\bol{x}_1}^{\bol{x}_s} \!dt\: \pp \dot{\xp} = \int_{\bol{x}_1}^{\bol{x}_s} \pp d\xp = \sum_i \int_{x_1^i}^{x_s^i} p_p^i d x_p^i\; .
\eeq
For concreteness, assume that the single peak in $\bol{x}_s$ is centered at $i = 0$.  Away from this peak, $x_s^i - x_1$ is small, and we expect that each integral $\int_{x_1^i}^{x_s^i} p_p^i d x_p^i$ is similarly small.  In fact, by linearizing Eq.\eref{eq:dyn} about $x_1$ and setting $\dot{x}^i = 0$, one immediately concludes that $x_s^i$ decays exponentially to $x_1$ as $|i|\rightarrow \infty$.  As long as the maximum over $t$ (or equivalently over \xp) of $|p_p^i(t)|$ approaches a finite value in this limit, the sum of integrals with respect to the $x_p^i$ is then bounded by a convergent geometric series, and $\Delta S_{1,s}$ must remain below a finite bound for all $M$.  It is not hard to see that $|p_p^i|$ in fact cannot diverge for large $|i|$:  Eq.\eref{eq:spec} is linear in $\bol{x_*}$ and approaches a screened diffusion equation for large $|i|$, where $x_p^i \rightarrow x_1$.  Thus, for large $|i|$, $x_*^i$ can be decomposed into an exponentially decaying part whose prefactor depends on nonlinear behavior in the peak near $i = 0$ and terms of order at most $x_p^i$.  Consequently, neither $x_*^i$ nor $p_p^i$ diverges for $|i| \rightarrow \infty$, and $\Delta S_{1,s}$ approaches a constant for large $M$, as asserted.

An interesting remark is that a similar construction for $\Delta S_{2,s}$ yields a dynamics analogous to Eq.~(\ref{eq:dyn}), but for trajectories between $x_2$ and $x_s$, that supports a propagating front solution for large enough systems.  This estimate thus gives an upper bound on $\Delta S_{2,s}$ that grows linearly with $M$, complementing our arguments on a lower bound in the main text and in the next section.

\vspace{15pt}

\subsection{Lower bound on $\Delta S_{2,s}$}
In this section, we elaborate on the arguments leading to the conclusion $\Delta S_{2,s}(M) \geq M \Delta_0$ for some positive $\Delta_0$, which we obtained in the main text. 

For states $\bol{x}_s$ and $\bol{x}_2$ connected by a solution $\bol{x}(t)$ of the deterministic kinetics\eref{eq:dyn}, one can rewrite $\Delta S_{2,s}$ as
\bea
-\Delta S_{2,s}(M) & = & S(\bol{x}_2)-S(\bol{x}_s) \nn
& = & \int_{\bol{x}_s}^{\bol{x}_2} \! dt\, \sum_{i=1}^{M} \frac{\partial S(\bol{x})}{\partial x^i} \dot{x}^i \nn
& = & \int_{\bol{x}_s}^{\bol{x}_2} \! dt\, \sum_{i=1}^{M} \biggl\{ \frac{\partial S(\bol{x})}{\partial x^i}  \nn
&& \times \left[w_+(x^i)-w_-(x^i)+ D\Delta_i (\bol{x})\right] \biggr\} \; . \nn
\label{eq:dete}
\eea
For large number of compartments $M$, the solution $\bol{x}(t)$ of the deterministic dynamical system
\beq
\dot x^i=w_+(x^i)-w_-(x^i)+ D\Delta_i (\bol{x})
\eeq
is a moving front (except in the immediate vicinity of the single peak in $\bol{x}_s$), and each term in the sum over $i$ should thus make the same contribution to the integral.  One might thus naturally be led to conclude that $\Delta S_{2,s}$ grows linearly with $M$ for large $M$.  One flaw in this reasoning is that we have not ruled out the possibility that $\partial S(\bol{x})/\partial x^i$ depends on $M$.  It will not do so as long as the dependence of the action $S(\bol{x})$ on the $x^i$ is sufficiently local, so that the derivative with respect to a given $x^i$ doesn't ``know'' how many compartments there are in total.  Our assumption in the main text that the solution to the full variational equations is a moving front can thus be replaced by the locality assumption that $\partial S(\bol{x})/\partial x^i$ is independent of $M$ for large $M$.  For this to be the case, it is sufficient that the second derivative $\frac{\partial^2 S(\bol{x})}{\partial x^i \partial x^j}$ have only short-ranged dependence on $i - j$:  $\frac{\partial^2 S(\bol{x})}{\partial x^i \partial x^j} < A \rme^{-C |i-j|}$ for some positive constants $A$ and $C$ and $|i - j|$ large enough.  This is known to be true in an equilibrium system, but remains to be proven completely rigorously for our nonequilibrium model.

\end{document}